\documentclass[journal]{IEEEtran}
\usepackage[normalem]{ulem} 
\usepackage{amsfonts}
\usepackage{amssymb}
\usepackage{amsmath}
\usepackage{setspace}
\usepackage{relsize}
\usepackage{color}
\usepackage{graphicx}
\usepackage{url}
\DeclareGraphicsExtensions{.pdf,.png,.jpg}
\usepackage{caption}
\usepackage{subcaption}

\begin{document}

\title{Early Signals from Volumetric DDoS Attacks: \\An Empirical Study}

\author{{\bf Michele Nogueira, Augusto Almeida Santos, Jos\'e M. F. Moura}
\thanks{M. Nogueira is with the Universidade Federal do Paran\'a, Caixa Postal 19072, Curitiba, PR, Brazil (michele@inf.ufpr.br).}
\thanks{A. A. Santos, and J. M. F. Moura are with the Dep. of Electrical and Computer Engineering, Carnegie Mellon University, Pittsburgh, PA 15213, USA (augusto.pt@gmail.com, moura@ece.cmu.edu). This work was partially supported by NSF grant CCF \#~1513936.} \thanks{{\bf Note:} This work has been submitted to the IEEE for possible publication. Copyright may be transferred without notice, after which this version may no longer be accessible.
}
}


\maketitle

\begin{abstract}

Distributed Denial of Service (DDoS) is a common type of Cybercrime. It can strongly damage a company reputation and increase its costs. Attackers improve continuously their strategies. They doubled the amount of unleashed communication requests in volume, size, and frequency in the last few years. This occurs against different hosts, causing resource exhaustion. Previous studies focused on detecting or mitigating ongoing DDoS attacks. Yet, addressing DDoS attacks when they are already in place may be too late. In this article, we consider network resilience by early prediction of attack trends. We show empirically the advantage of using non-parametric leading indicators for early prediction of volumetric DDoS attacks. We report promising results over a real dataset from CAIDA. Our results raise new questions and opportunities for further research in early predicting trends of DDoS attacks.

\end{abstract}



\section{Introduction}

Distributed Denial of Service (DDoS) attack is one of the major threats for Cyberspace. Attackers have intensified malicious DDoS in speed, volume, and sophistication~\cite{Man:2015}. Their main goal lies in overloading links or servers. DDoS attacks can reach volumes of hundreds of Gigabits per second
~\cite{Man:2015,Zar:2013}.
For sophistication, attackers take advantage of advances in computing and communication technologies, by controlling geographically-distributed infected machines (a.k.a., `bots'). These machines together can produce simultaneously thousands of requests from different Internet connections~\cite{Far:2016}, making it extremely difficult to detect and prevent the attack.  

Detecting, controlling, and restraining volumetric DDoS attacks are challenging. Attackers constantly enhance their techniques and produce higher and longer volumes of traffic congestion. 
They tend to follow the cycle: infect, coordinate, and attack~\cite{Zar:2013}. Attackers infect machines with malwares. These malwares coordinate themselves to define the target, and then they launch the attack. Bots can learn about the network behavior, -- based on e.g. the current response latency on servers, --  and maliciously use this knowledge. Once coordinated, bots act fast against the target~\cite{San:2016}. They launch a tremendous amount of requests or increase inadvertently the size of network packets. These actions lead to overloaded links or servers.

Several studies have primarily concentrated on parametrically detecting DDoS attacks, e.g.,~\cite{Tab:2016,Tsa:2010}. But, attack detection approaches are currently limited. Attack behavior changes constantly and quickly, making difficult to keep detection systems updated. Hence, mitigation approaches have sought to reduce the impact of ongoing volumetric DDoS attacks. These approaches offer flexibility. They apply recent network paradigms, such as software-defined networking (SDN) and network function virtualization (NFV)~\cite{Fay:2015}. Yet, mitigation approaches focus on ongoing attacks, but these can be harder to control. In this work, we advocate for the non-parametric predicting of trends of DDoS attacks even before the onsetting of the attack. The non-parametric prediction of DDoS attack trends complements recent advances in network security. Trends can serve as a reference for mitigation and other defense actions.

Ours is a first work that explores {\em leading indicators} for predicting trends of DDoS attacks. We show empirically the potential of these indicators to provide early warnings of DDoS attacks. Leading indicators consist of general characteristics observed in advance of disruptive changes~\cite{Mar:2009}. Such characteristics are tied to metastability phenomena~\cite{Bov:2015}. We show in this article how to use them to predict changes. By disruptive change, we mean a critical transition in the network state, like for instance, when the attack suddenly leads the network to an unexpected state. This new state can incur large costs as restoration to the previous state is difficult or impossible.

We explore empirically a set of classical leading indicators~\cite{Mar:2009}: return rate, autocorrelation, coefficient of variance, and skewness. In contrast with other works, we employ these leading indicators in the context of network security. We show that the indicators can be used to predict trends of volumetric DDoS attacks. We calculate the leading indicators for time series. The time series represent the network load by packet size as a function of time. We extract the time series from tcpdump traces of a dataset from the {\em Center for Applied Internet Data Analysis} (CAIDA)\footnote{The CAIDA UCSD "DDoS Attack 2007" Dataset: \url{http://www.caida.org/data/passive/ddos-20070804_dataset.xml}}. The traces are from a real DDoS attack. They include only attack traffic to the victim and responses from the victim. As much as possible, CAIDA has removed non-attack traffic. We report promising results from before and during the attack launching time. We evaluate changes of the leading indicators. Our results discover a set of behaviors that are expected and can be used to predict trends of a disruptive change in the network state. Also, they suggest a set of relevant questions for future research.

This article proceeds as follows. Section~\ref{sec:attacks} overviews volumetric flooding-based DDoS attacks. Section~\ref{sec:indicators} describes metastability and the leading indicators that we employ. Section~\ref{sec:eval} details the method followed in our analyses and Section~\ref{sec:results} reports the results. Section~\ref{sec:relworks} discusses related work. Finally, Section~\ref{sec:conclusion} concludes the article.

\section{DDoS Attacks}
\label{sec:attacks}

A basic DDoS attack consists in flooding a single server with thousands of requests from bots~\cite{Man:2015,Zar:2013}. Hence, the server becomes completely overloaded and can no longer respond to legitimate user requests. The bots are geographically distributed. They depend on different types of Internet connections, making it difficult to control attacks. These machines coordinate themselves in networks (botnets) compounding their effect.

DDoS attacks have exploited different protocols such as TCP, HTTP, and DNS to cause significant damage to the Internet. Lately, attackers have even exploited social networks and cloud systems to boost DDoS effects~\cite{Man:2015}. Today, the use of mobile devices presents new vulnerabilities. These devices allow attackers to take advantage of device-to-device wireless communication. This type of communication can mask malicious code propagation. Then, attackers can recruit a larger number of bots and increase DDoS attack traffic. These devices also produce a dynamic and adaptive behavior for the attack, making it harder to design an effective defense. 

The volumetric flooding-based DDoS attack is a specific type of DDoS. It acts by increasing the number of requests or packet size. Increasing the number of requests aims at exhausting server processing, whereas increasing the packet size has the goal of overloading network resources, such as bandwidth. In general, attackers increase the number of requests and packet size simultaneously. Given its simplicity, this type of DDoS attack is becoming increasingly more significant~\cite{Man:2015}. Furthermore, it has compromised different companies' websites, resulting often in significant financial losses. 

This article focuses on volumetric flooding-based DDoS attacks. These attacks are effortless to launch, but they can have significant power to disrupt the entire state of the network. We abstract two features for analysis: the number of requests and packet size. Then, we investigate empirically the applicability of recently discovered behavior exhibited by classical non-parametric leading indicators. We will show that these indicators can help predicting trends of DDoS attacks even before they effectively start.  

\section{Leading Indicators for DDoS Attacks}
\label{sec:indicators}

We show how generic symptoms may indicate a volumetric flooding-based DDoS attack. To measure such symptoms, we exploit a set of four classical statistical leading indicators: return rate, autocorrelation, coefficient of variance, and skewness~\cite{Mar:2009}. In the next subsections, we introduce the concept of metastability and the leading indicators for such phenomena. For details about leading indicators, see~\cite{Mar:2009}. Leading indicators are well known in Statistics~\cite{Mar:2009}. The novelty here lies in investigating their applicability to predict trends of a disruptive transition. Their applicability is based on certain properties found in metastability. More specifically, our work uses leading indicators to predict trends of volumetric flooding-based DDoS attacks. 

\subsection{Metastability}

A system exhibits metastability when it tends to spend a long time in a particular intermediate state~$A$, a.k.a metastable state (or metastate), and then eventually it drifts swiftly to another state~$B$, where it remains now for a long time (or indefinitely). More precisely, the dynamics of the system can be partitioned into three (or more) distinct time-scales: i) long-time at metastate~$A$; ii) rapid transition to a different state~$B$; iii) staying around~$B$ for a long (possibly an infinite amount of) time. Characterizing metastability requires determining the metastates, the (random) times that the system spends at each metastate, and how fast is the transition among them. The literature on the subject is dominated by empirical studies, as formal analysis is hard. Observing empirically macroscopic metastability is far easier than studying it analytically from the microscopic interactions among agents. For some analytical approaches on the mathematical formalism of metastability we refer the reader to~\cite{Metastability}.

\begin{figure} [hbt]
\begin{center}\includegraphics[scale=0.4]{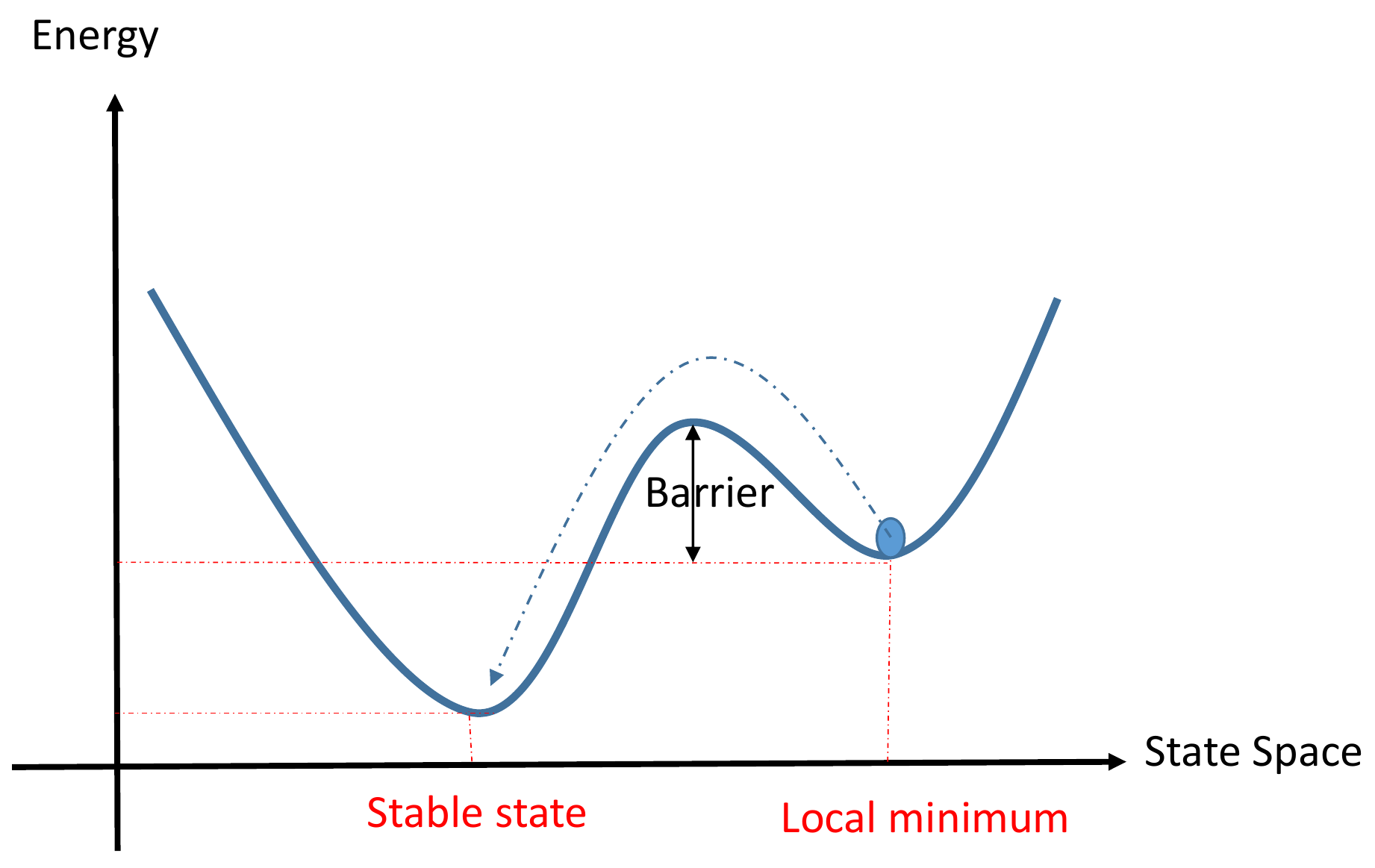}
\caption{Free energy associated with the state of a system. The system tends to drift to low energy states and may be trapped in local minima regions.}\label{fig:freenergy}
\end{center}
\end{figure}

Metastability often arises in systems that exhibit two or more possible stable states. For instance, consider the framework of diffusion of two (for simplicity, but in practice  usually more) opinions~$A$ and~$B$ in a social network of individuals, where every individual exerts some influence upon their peers' opinions and is influenced by theirs as well. There are at least two stable states: everyone adopts opinion~$A$ or everyone adopts opinion~$B$. The system may swing swiftly between the two states depending on exogenous perturbations. We give an intuitive interpretation of the stable states of a dynamical system now from an energetic point of view. Each state corresponds to a certain level of energy, and the stable states are those local (or global) minima of an energy function as exemplified in Fig.~\ref{fig:freenergy}, often referred to as free energy states. A stochastic system undergoes perturbations about a stable state, and if the deviations are large enough, it will overcome a barrier of energy and will drift quickly to another stable state. A relevant measure of the time it takes for a system to move from one to another stable state  is the return rate that we discuss next.

As a metastable system undergoes a fast transition -- possibly, to an undesirable state (e.g., blackout in a power grid, sudden compromising of IoT services, cardiac arrest) -- it is reasonable to respect \emph{early-warning signal} precursors before the transition occurs, which in turn can help with controlling it. More formally, these precursors may be used to flag when the probability of transition becomes unusually high. 
It is an important and subtle aspect of the problem to characterize such macroscopic observables. Depending on the application, several heuristic statistics are used in the literature to forecast a transition beyond the four leading indicators referred to before~\cite{Wang}. 
As a general rule, in interacting particle systems~\cite{Metastability} (the framework that formally studies metastability), there are no broad universal approaches and predictive statistics are usually crafted targeting each scenario. One exception to the rule is the seemingly universal usefulness of the return rate. It measures the irreversibility of a transition, i.e., if the system is drifting from a metastable state to another.

\subsection{Leading indicators}
\label{sub:leading}

In this subsection, we explain the four leading indicators -- return rate, autocorrelation, coefficient of variance, and skewness -- from the perspective of a computer network under a volumetric DDoS attack. Before presenting each one of the leading indicators, we associate the metastability phenomenon with a computer network under an attack. Taking Fig.~\ref{fig:freenergy} as basis, imagine the network in a metastable state, as initially indicated by the blue circle. The DDoS attack can produce a disturbance in the network state. Disturbances can push the state towards values that are close to conquering the barrier between the two metastates. Small disturbances may cause changes in the network state that may not be enough to make it to overcome the barrier. If the disturbance is big enough to potentially push the state across the barrier, this means that the network state experiences a critical transition. 

In this article, we analyze network states based on the network packet sizes in a time window. Hence, a network state is characterized by the behavior of a set of network packet sizes during a time window. Each packet size is observed from the perspective of the packet destination, and it is indexed per time unit $t$ (a time series). We consider that each packet is collected sequentially, hence each packet is labeled by a unique timestamp. For each time window, we organize the observables in a matrix, in which the cell contains the packet size per time unit (matrix column) and packet destination (matrix row). 
We have chosen to observe packet size because of the general behavior of a volumetric DDoS attack. We know that packet size is a network feature susceptible to attacker manipulation. Attackers can easily produce fake network packets containing bigger sizes in order to overload the network bandwidth. Hence, we take this feature to analyze and investigate the network behavior and patterns over time. 

Now, we start to explain the four leading indicators. The first one is the rate of return or recovery to equilibrium ({\bf return rate}). 
In technical terms, it is calculated by the {\em dominant eigenvalues} from the matrix composed of the observables during a time window. The dominant eigenvalues characterize the rate of change around a metastable state. It can indicate the proximity of a critical transition. Particularly, it can indicate the irreversibility of a transition, i.e., if the network state is drifting from a metastable state to another, overpassing the barrier. The higher the return rate, the faster the network recovers from small disturbances around its current state and, consequently, we say that the network has a high resilience to change. The return rate is reduced when the network approaches a critical transition, decreasing smoothly to zero as the disruptive transition approaches~\cite{Mar:2009}. The smooth reduction of the return rate close to a critical transition is called {\em Critical Slowing Down} (CSD), a term coined from dynamical systems theory.

Regarding the second leading indicator, {\bf autocorrelation} is a metric employed to evaluate correlation between observables. It measures how much the network states become increasingly similar between consecutive observations. Scheffer et. al~\cite{Mar:2009} observed that the autocorrelation in time series increases when critical transitions approach. Disruptive transitions tend to increase the correlation at low lags (also known as `short-term memory') between observables of the time series. To be specific, we calculate the autocorrelation at-lag-1, i.e., the correlation of the time series to itself shifted one time-step back. In Eq.~\ref{eq:lag1}, the variables $z_t$ and $z_{t+1}$ represent two consecutive observables at times $t$ and ${t+1}$, respectively, and $\mu$ the mean in a given time window, and $\sigma$ is the variance of the variable $z_t$.

\begin{equation}
\rho_1 = \frac{\mathbb{E}[(z_t - \mu)(z_{t+1} - \mu)]}{\sigma_{z}^2}
\label{eq:lag1}
\end{equation}

The third leading indicator is the {\bf coefficient of variation}. It is a statistical measure that indicates the level of variance in a time series. It is calculated as $CV = \frac{SD}{\mu}$, where SD is the empirical standard deviation calculated by $SD = {\sqrt{\frac{1}{n-1}\sum\limits_{t=1}^{n}(z_t - \mu)^2}}$. From the literature~\cite{Mar:2009}, critical transitions and the CSD phenomenon increase the variance in a time series. Hence, together to the other leading indicators, the coefficient of variance can assist in predicting trends of a critical transition, in this case a volumetric DDoS attack.

The fourth leading indicator is {\bf skewness}, a well-known statistic measure that indicates the asymmetry in the probability distribution of observables about its mean. The skewness value can be positive or negative, or even undefined. In simplistic terms, the rise in skewness means that the distribution of the observables will become asymmetric. Skewness is calculated as illustrated in Eq.~\ref{eq:gamma}. Like variance, skewness can also increase because of a critical transition, meaning that the asymmetry in the observables increases. This happens because the dynamics close to the barrier become slow. Scheffer et. al~\cite{Mar:2009} have observed a rise in the skewness in different types of time series, such as time series from Ecology or Finance systems, when they are close to a critical transition. In this work, we employ skewness as a leading indicator and it is analyzed together with the other three.

\begin{equation}
\gamma = \frac{\frac{1}{n}\sum\limits_{t=1}^{n}(z_t-\mu)^3}{\sqrt
{\frac{1}{n}\sum\limits_{t=1}^{n}(z_t - \mu)^2}}
\label{eq:gamma}
\end{equation}

The return rate and the three statistical measures 
are employed as leading indicators about trends of DDoS attacks. The individual analysis of a single leading indicator should not be taken as basis for conclusions, in order to reduce false positives or false negatives. Their analysis should be performed together.  Scheffer et. al~\cite{Mar:2009} have demonstrated in other domains -- e.g., mainly, Ecology and Finance -- that this set of leading indicators presents specific behavior in the imminence of a critical transition. In summary, they observed that in the imminence of a disruptive change $i)$ the return rate decreases; $ii)$ the autocorrelation at-lag-1 increases; $iii)$ the coefficient of variance increases; and $iv)$ the skewness increases. In the next section, we employ this set of leading indicators for the early prediction of volumetric DDoS attacks. To the best of our knowledge, this is the first time this theory is applied to computer networks under DDoS attacks. We calculate and compare the values for these leading indicators over one-hour trace from a real DDoS attack.  

\section{Attack Analysis}
\label{sec:eval}

We consider the CAIDA ``DDoS Attack 2007" 
dataset as basis for our analysis. The dataset contains anonymized traces from a real volumetric DDoS attack (TCP-Like SYN flooding). The attack occurred on August 4, 2007. It attempts to block all access to a targeted server. Hence, bots consume the computing resources of the server. Also, they waste the network bandwidth connecting the server to the Internet. In particular, the attack generates a large amount of regular and oversized SYN packets. As observed from the traces, SYN packet sizes vary from 60 to 1500 bytes.

The total size of the dataset is 21 GB and it is about one hour duration (20:50:08 UTC to 21:56:16 UTC). The attack itself starts around 21:13, as labeled by CAIDA, when the network load increased rapidly (within a few minutes) from about 200 kbits/s to about 80 Mbits/s. The one-hour trace is split in one-minute (0 to 59.99 seconds) time series. The set of time series was extracted from pcap files by tcpdump filters. The duration of the time series has been chosen due to the large amount of data and just to simplify the  calculation of the leading indicators. The size of each time series file varies from a few Megabytes to Gigabytes. 
The traces include only attack traffic {\em to the victim} and {\em responses from the victim}. CAIDA has removed as much as possible non-attack traffic.
There are three different moments in the attack timeline. We call these moments: preparation (Phase 1), attack kickoff, and attack time (Phase 2). In this work, we concentrate in the {\bf preparation phase} and in the {\bf attack kickoff}, since our goal is to indicate trends of attack imminence. 

In order to estimate leading indicators and analyze their behavior, time series are composed of a relative time (from 0 to 59.99 seconds) and packet size. The goal lies in employing relevant but simple features in order to predict the attack imminence. Features such as packet generation frequency and packet size can together strongly characterize a volumetric DDoS attack. High frequencies in packet generation are in general associated with DDoS; while increases in packet size indicate that the attacker is trying to cause network saturation. According to the network protocol, packet sizes should be less than 250 bytes. However, we observe in the traces, packet sizes reaching 1500 bytes, mostly when the network is under attack.

\begin{figure}[!h]
\centering
\vspace{-0.3cm}
\includegraphics[width=.5\textwidth]{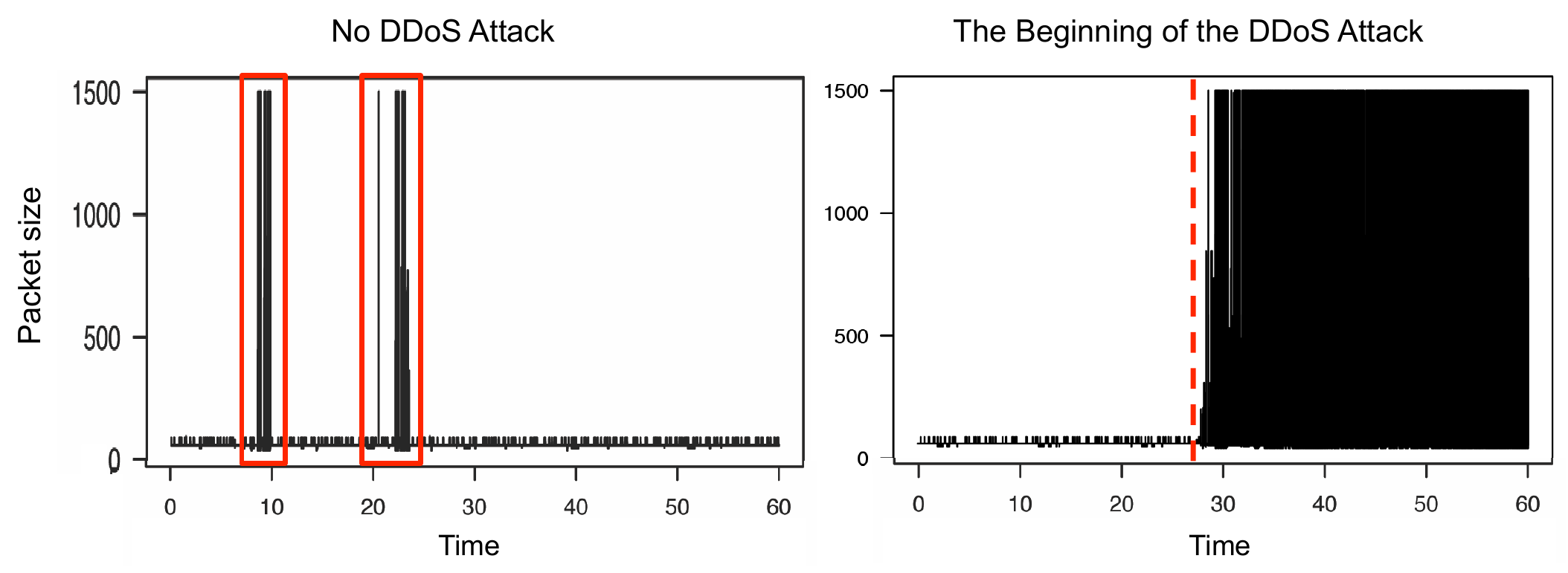}
\caption{Time Series: Packet size vs Relative time. Examples from the preparation phase (at 20:55)(left) and kickoff time (at 21:13)(right).}
\label{fig:timeseriescombo}
\vspace{-0.4cm}
\end{figure}

Fig.~\ref{fig:timeseriescombo} shows two of the time series extracted from the dataset. They are from different moments in the attack timeline. The first is from the period before the time indicated by CAIDA as the beginning of the attack. The second is from the attack kickoff. 

From these time series, one can observe a fast increase in the packet size (Fig.~\ref{fig:timeseriescombo}, right). While in the first time series (Fig.~\ref{fig:timeseriescombo}, left) the majority of the packets are of 60 bytes, in the second time series (Fig.~\ref{fig:timeseriescombo}, right) one can see the transition (at the red dashed vertical line) from a period in which packets are of 60 bytes to a period in which the majority of the packets are 1500 bytes. The attack starts at the relative time 30 (i.e., 21:13 UTC) of the second time series (Fig.~\ref{fig:timeseriescombo}, right) and lasts for many minutes (not presented in the figure). Another interesting aspect lies in the two picks (highlighted by the red boxes) presented in the first time series (Fig.~\ref{fig:timeseriescombo}, left). Despite the fact that this time series is from Phase 1, one can observe few packets of 1500 bytes. We have observed this behavior in other analyzed time series from Phase 1. We suspect that they correspond to bots preparing and testing themselves, a few minutes before launching the attack.        

\begin{figure}[t]
\begin{subfigure}{.5\textwidth}
\hspace{1.1cm}
  \includegraphics[width=0.70\linewidth]{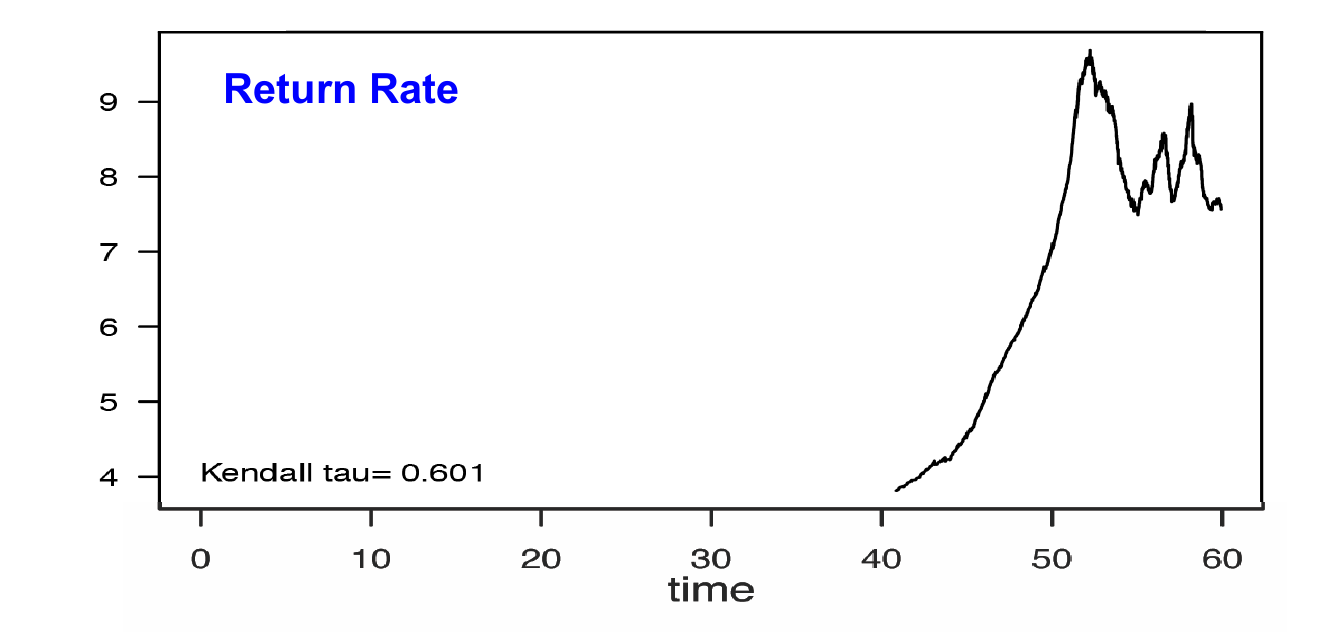}
\label{fig:RR}
\end{subfigure}\\
\begin{subfigure}{.5\textwidth}
 \centering
\includegraphics[width=0.65\linewidth]{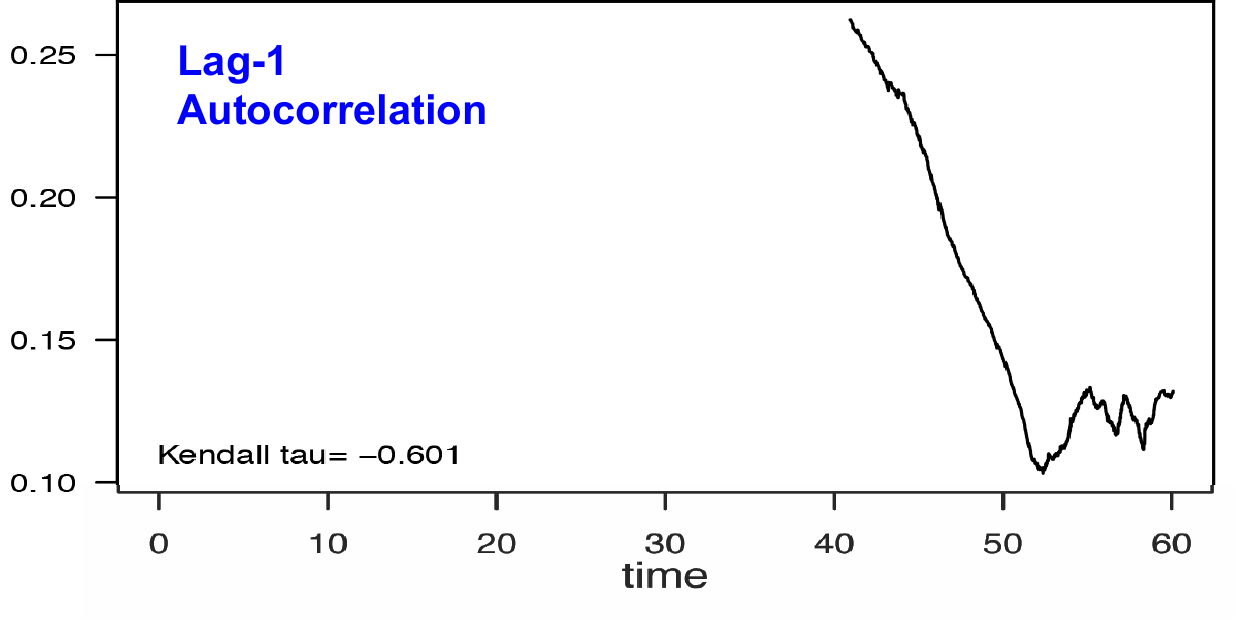}
\label{fig:ACF}
\end{subfigure}\\
\begin{subfigure}{.5\textwidth}
\centering
\hspace{0.1cm}
\includegraphics[width=0.66\linewidth]{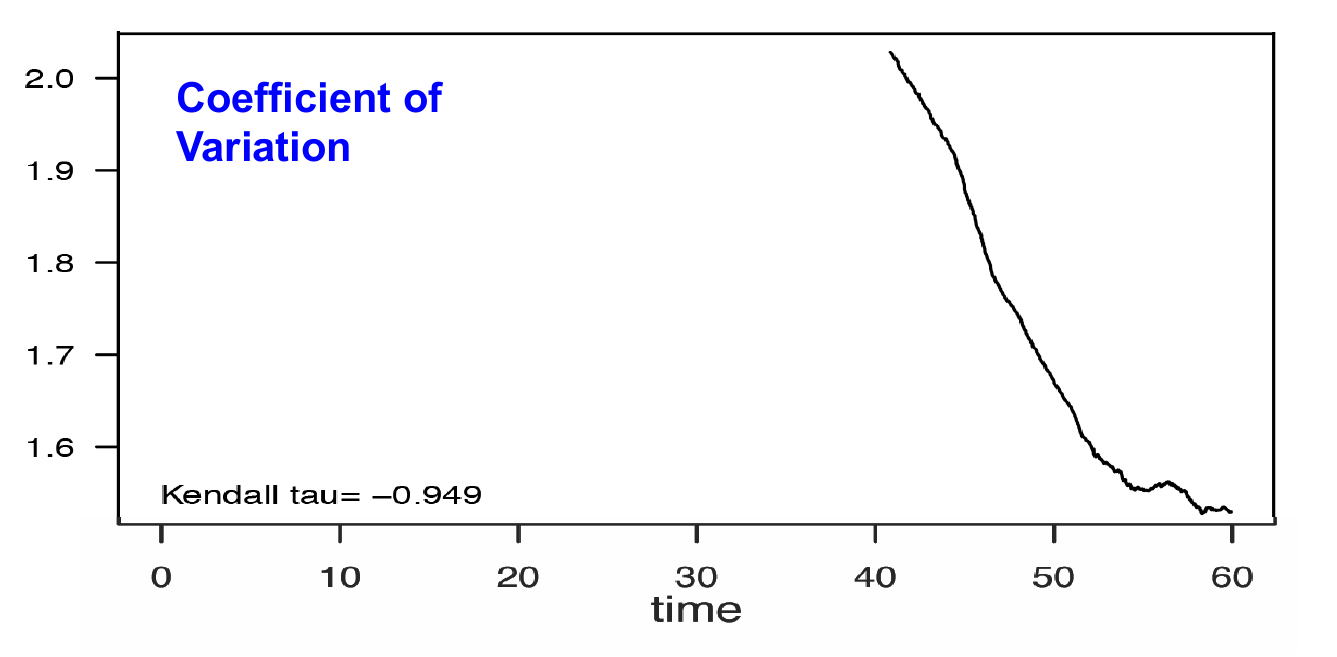}
\label{fig:CV}
\end{subfigure}\\
\begin{subfigure}{.5\textwidth}
\centering
\includegraphics[width=0.65\linewidth]{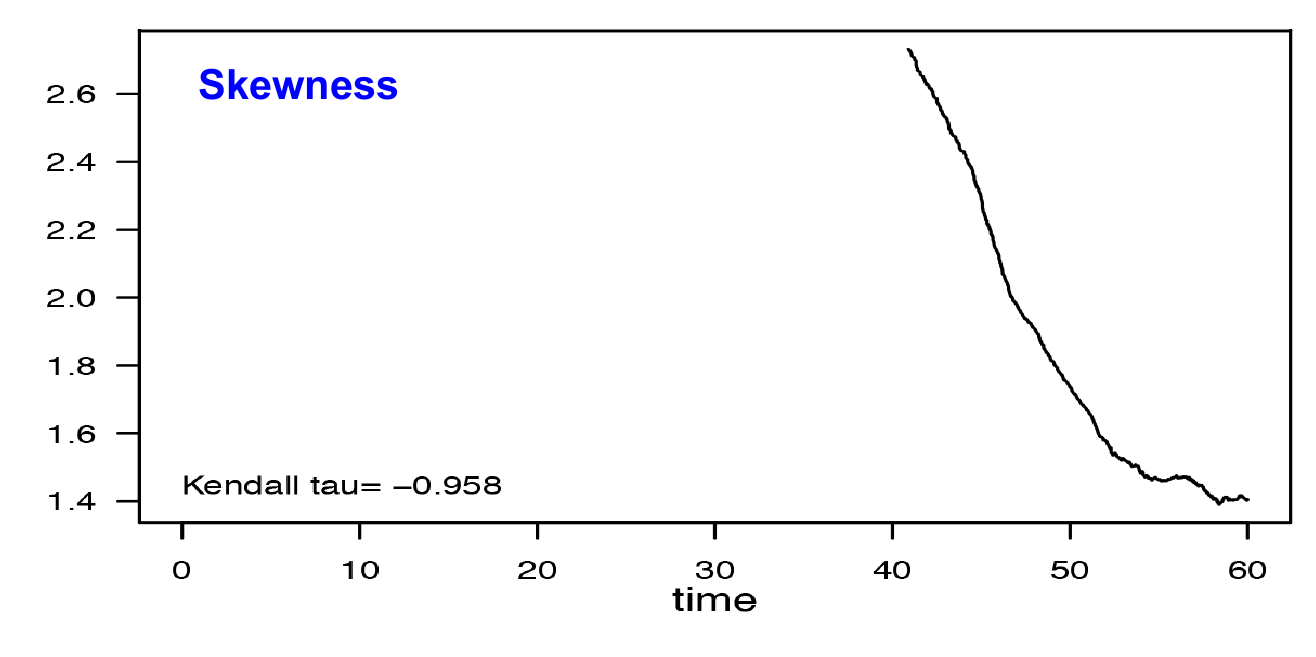}
\label{fig:Skew}
\end{subfigure}
\caption{Leading Indicators in attack kickoff (at 21:13). Note: leading indicators {\em do not} follow the behavior expected for a critical transition. }
\label{fig:kickoff}
\vspace{-0.4cm}
\end{figure}

\section{Results}
\label{sec:results}

This section presents the results from this study in applying the set of leading indicators to the CAIDA dataset. We organize the section around two main questions: (1) ``Could we identify trends of the attack at its launching time?" and (2) ``Could we explore leading indicators in the preparation phase?" These questions guide our analyses and help us to draw conclusions. 

\subsection*{Could we identify trends of the attack at its launching time?}
 
{\em No.} We estimated and analyzed the leading indicators at the attack kickoff. We expected to observe the characteristic behavior of a disruptive transition at that moment. However, as one can observe in Fig.~\ref{fig:kickoff}, the behavior of the leading indicators is different from those reported in the literature as characterizing a disruptive transition (see Subsection~\ref{sub:leading}). The return rate tends to increase, instead of decreasing, and then it presents some variation. The lag-1 autocorrelation decreases, instead of increasing. The coefficient of variation decreases, instead of increasing, as well as the skewness distribution. Hence, the behavior for the leading indicators during the attack kickoff does not reflect an expected behavior for the imminence of a disruptive transition. 
We could not identify trends of the attack at its launching time using these leading indicators. 

\begin{figure}[t]
\begin{subfigure}{.5\textwidth}
\hspace{1.1cm}
  \includegraphics[width=0.68\linewidth]{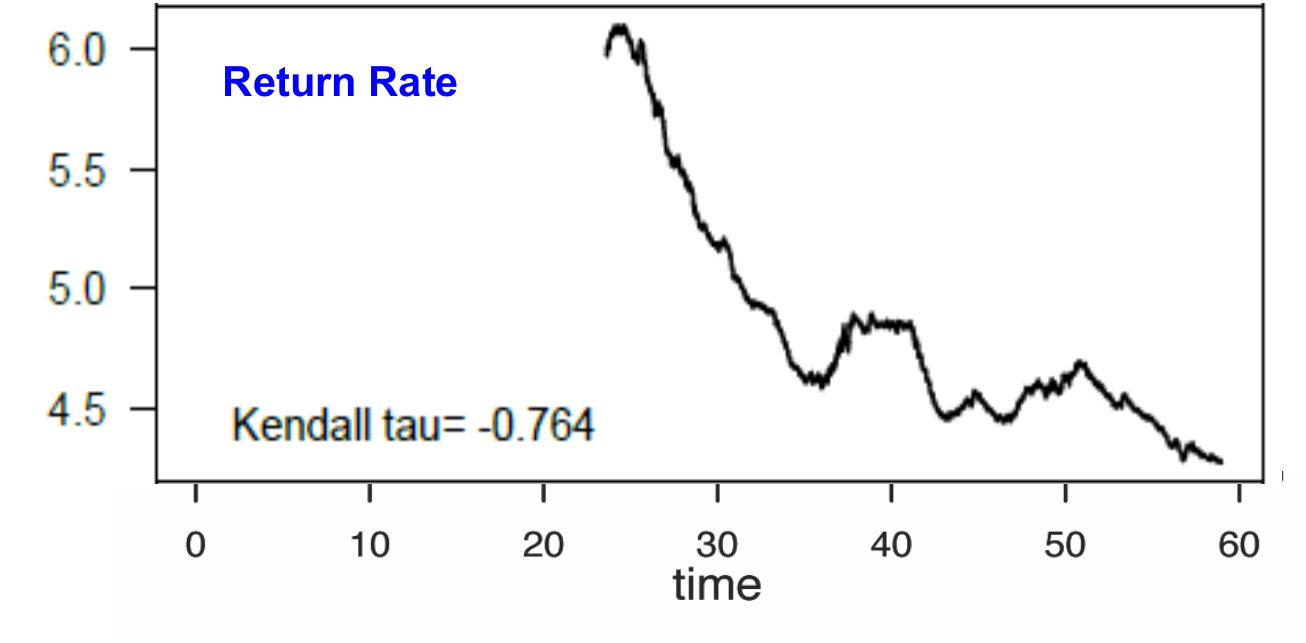}
\label{fig:RR}
\end{subfigure}\\
\begin{subfigure}{.5\textwidth}
 \centering
\includegraphics[width=0.65\linewidth]{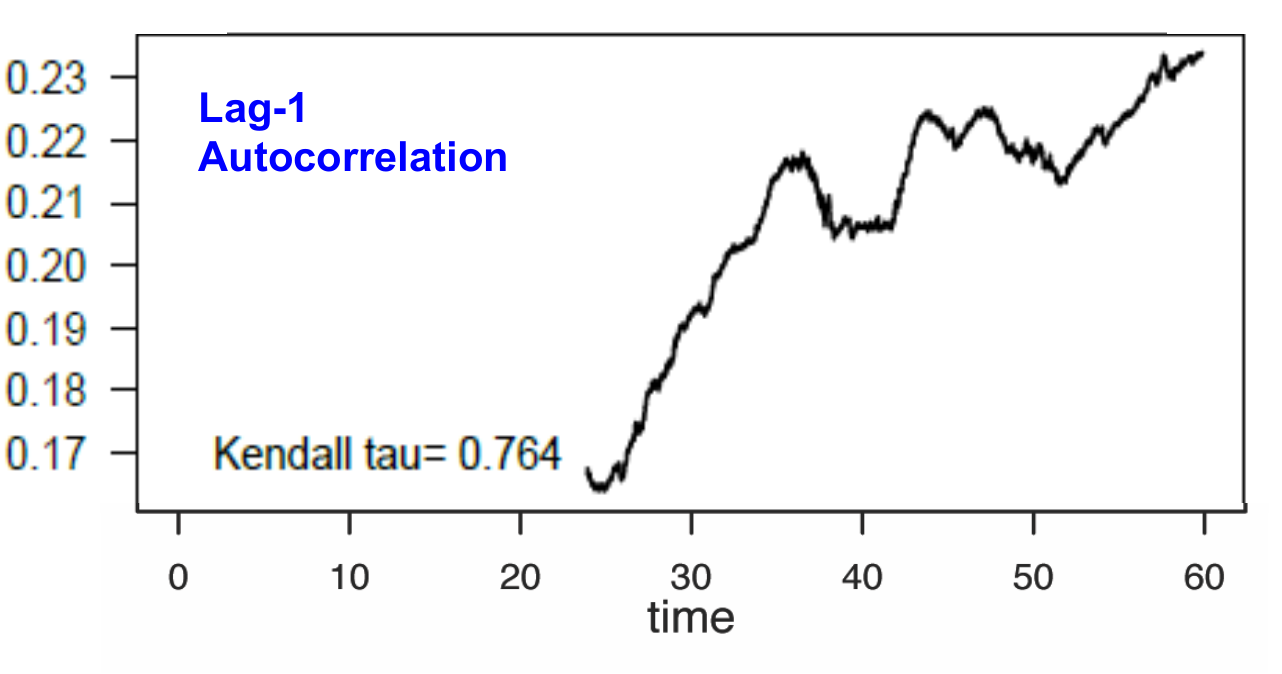}
\label{fig:ACF}
\end{subfigure}\\
\begin{subfigure}{.5\textwidth}
\centering
\includegraphics[width=0.65\linewidth]{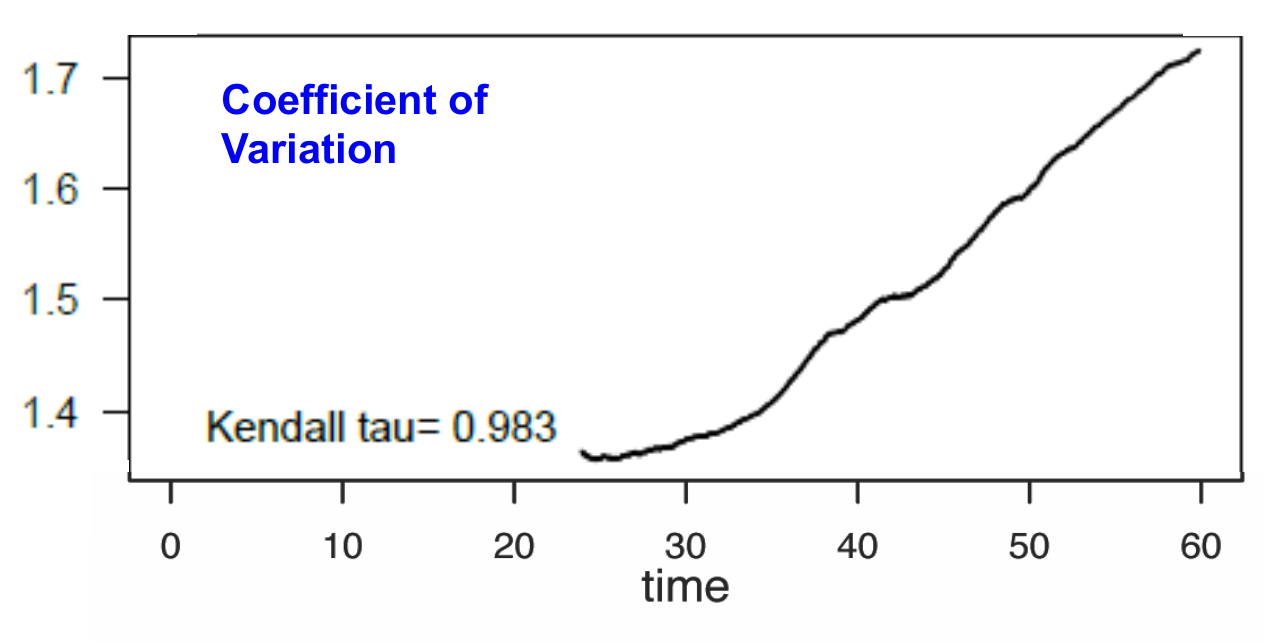}
\label{fig:CV}
\end{subfigure}\\
\begin{subfigure}{.5\textwidth}
\centering
\includegraphics[width=0.66\linewidth]{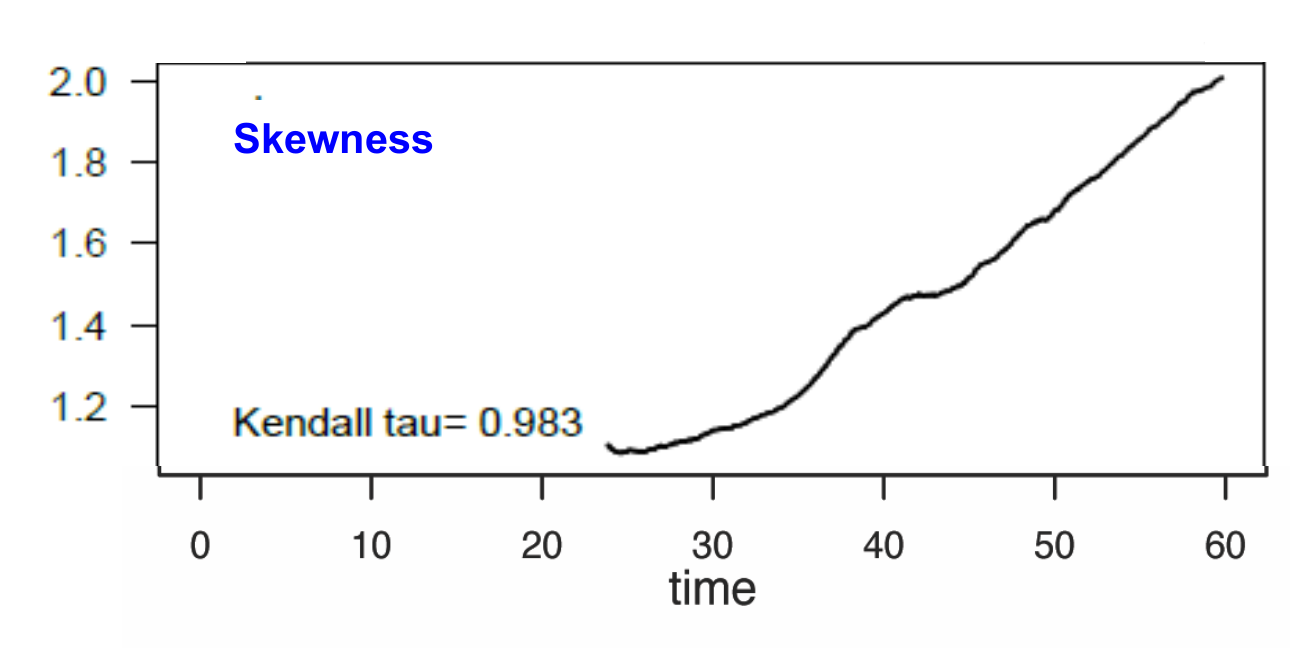}
\label{fig:Skew}
\end{subfigure}
\caption{Leading indicators in the minute 2 (at 20:51) of the Dataset (Phase 1). Note: the characteristic behavior of a critical transition approximation - {\em decrease} in return rate, and {\em increase} in autocorrelation, coefficient of variation, and skewness.}
\label{fig:prepphase}
\end{figure}

\subsection*{Could we explore leading indicators in the preparation phase?}

{\em Yes.} We examined the leading indicators in different time series from the preparation phase. Surprisingly, we identified the disruptive transition behavior, precisely, at the second and third minutes of the dataset, i.e., 20:51 and 20:52, respectively. At these minutes, it is already possible to identify suspicious behavior identified by the leading indicators. A few other minutes of the dataset in the preparation phase also present this trending behavior. We do not show all of them in this work due to limitations in terms of space and number of figures. However,  an illustration of the leading indicators is presented in Fig.~\ref{fig:prepphase}. One can observe a decrease in the return rate as well as increase of the lag-1 autocorrelation, coefficient of variation and skewness, as expected from the literature characterizing a disruptive transition. 

\section{Related Work}
\label{sec:relworks}

There has been several papers published in early warning systems in the last few years. 
In~\cite{Ram:2016}, Ramaki and Atani presented a survey of architectures and techniques for early warning threats in Information Technology (IT). The authors classify the early warning systems (EWS) in commercial or under research and development. They also point out to a set of current challenges, such as data collection, data correlation, and post-event data correlation~\cite{Ram:2016}. The authors reinforced the need for designing proactive solutions to predict threats and attacks before they occur in the system by using data analytics.

Different studies have particularly tackled the problem of developing early warning techniques for DDoS attacks~\cite{Tsa:2010,Xia:2006,Kor:2016}. In~\cite{Xia:2006}, Xiao, Chen and He proposed a cooperative system to produce warning signals. The system is based on a Bloom filter technique. The authors' goal is to reduce storage and computational resources consumption. In~\cite{Tsa:2010}, Tsai, Chang and Huang presented a multilayer system based on time delay neural networks. Their system is cooperative with each device in the network monitoring its neighbors. At a certain point, the device sends the collected data to an expert module. The module then analyzes all collected data and attempts to match the received data with known DDoS patterns. 
    
In~\cite{Kor:2016}, Korczy\'nski, Hamieh, Huh, Holm, Rajagopalan and Fefferman presented a cooperative and self-organized anomaly detection system inspired by colonies of honey bees. Their goal was to provide dynamic thresholds to detect  anomalous patterns in network traffic. Also, they intended to improve early intrusion detection in order to assist in mitigation of attacks. 

These works advanced the literature of early warning systems against DDoS attacks. However, in general, they address the problem under the perspective of designing a full system. Our work here complements theirs, because we aim at defining generic indicators to highlight the possible imminence of attacks. These indicators can serve as basis for early warning systems, assisting them in identifying trends of DDoS attacks before they are fully in place.

\section{Conclusion}
\label{sec:conclusion}

This article investigates the potential of leading indicators to early identify trends of Distributed Denial of Service (DDoS) attacks. Leading indicators are a set of generic characteristics based on mathematical properties related to metastability phenomena observed when approaching a disruption, i.e., a DDoS attack. This work gives insight about their potential to predict DDoS attacks trends before their kickoff. Our results with real data showed that they exhibit specific behaviors before the onsetting of an attack.  
These leading indicators, i.e., return rate, autocorrelation, coefficient of variation, and skewness, exhibit the characteristic behavior of a critical transition 22 minutes before the attack kickoff. Our study also shows that the leading indicators did not present this behavior at kickoff time. 
These results raise new questions and opportunities for further investigation regarding early prediction of DDoS attack trends and how to prevent them efficiently in the Internet. Further work will focus on investigating different attack features, that can serve as basis to evaluate the leading indicators as well as their application to other datasets. 

\section*{Acknowledgments}

Support for CAIDA's Internet Traces is provided by the National Science Foundation, the US Department of Homeland Security, and CAIDA Members. This work was partially supported by CNPq, Funda\c{c}\~ao Arauc\'aria and CAPES -- Strategic Research Program -- Grant 99999.000404/2016-00 and by NSF grant \#CCF-1513936.

\bibliographystyle{IEEEtran}
\bibliography{refs,biblio}

\end{document}